\documentclass[aps, twocolumn,superscriptaddress,amsmath,amssymb,showpacs]{revtex4}
\usepackage{amsmath,amsfonts,amssymb}
\usepackage{graphicx,subfigure}
\usepackage{epsfig}

\begin{document}
\title{Bright gamma-rays from betatron resonance acceleration in near critical density plasma
}

\author{B. Liu}
\affiliation{Institute of Applied Physics and Computational
Mathematics, Beijing, China, 100088}

\author{H. Y. Wang}
\affiliation{Key Laboratory of HEDP of the Ministry of Education,
CAPT,and State Key Laboratory of Nuclear Physics and Technology,
Peking University, Beijing, China, 100871}

\author{D. Wu}
\affiliation{Key Laboratory of HEDP of the Ministry of Education,
CAPT,and State Key Laboratory of Nuclear Physics and Technology,
Peking University, Beijing, China, 100871}

\author{J. Liu}
\affiliation{Institute of Applied Physics and Computational
Mathematics, Beijing, China, 100088}

\author{C.E.Chen}
\affiliation{Key Laboratory of HEDP of
the Ministry of Education, CAPT,and State Key Laboratory of Nuclear
Physics and Technology, Peking University, Beijing, China, 100871}
\author{X. Q. Yan}
\email[]{X.Yan@pku.edu.cn}
\affiliation{Key Laboratory of HEDP of
the Ministry of Education, CAPT,and State Key Laboratory of Nuclear
Physics and Technology, Peking University, Beijing, China, 100871}

\author{X. T. He}
\email[]{xthe@iapcm.ac.cn}
\affiliation{Institute of Applied Physics and Computational
Mathematics, Beijing, China, 100088}
\affiliation{Key Laboratory of HEDP of
the Ministry of Education, CAPT,and State Key Laboratory of Nuclear
Physics and Technology, Peking University, Beijing, China,
100871}

\begin{abstract}
We show that electron betatron resonance acceleration by
an ultra-intense ultra-short laser pulse in a near critical density plasma
 works as a high-brightness gamma-ray source.
Compared with laser plasma X-ray sources
in under-dense plasma,
near critical density plasma provides three benefits for electron radiation:
more radiation electrons, larger transverse amplitude, and higher betatron
oscillation frequency.
Three-dimensional particle-in-cell simulations show that,
by using a 7.4J laser pulse, 8.3mJ radiation with
critical  photon energy 1MeV is emitted.
The critical photon energy $E_c$
increases with the  incident laser energy 
$W_I$ as $E_c \propto W_I^{1.5}$, and the corresponding photon number is
proportional to $W_I$.
A simple analytical synchrotron-like radiation model is built,
which can explain the simulation results.
\end{abstract}


\maketitle

High-brightness high-speed X-ray pulses have become  
powerful tools for a wide variety of scientific applications in
 physics, chemistry, biology, and material science, etc.
X-ray pulses can be generated when  relativistic electrons experience
transverse oscillations.
The traditional X-ray sources, such as synchrotron radiation sources and Compton scattering sources,
are usually based on the conventional particle accelerators,
which are very large and expensive.
Recently, with the rapid development of laser-driven acceleration technology,
all optical X-ray sources, which are compact and cost-effective, attract many interests
\cite{rmp_x-ray}.

When a relativistic electron experiences transverse oscillation,
with Lorentz factor $\gamma$, transverse velocity $v_{\perp}$,
and transverse oscillating frequency $\omega_{\beta}$,
X-ray pulse will be radiated, with
critical photon energy \cite{textbook}
 \begin{equation}
	E_c \sim  \hbar  \omega_{\beta} \gamma^3 v_{\perp}/c,
\end{equation}
radiation power 
$P \sim 2 \alpha E_c  \omega_{\beta} \gamma v_{\perp}/(3c)$,
and confined in a narrow
angle $\Delta \theta \sim 1/\gamma$ along the electron motion direction,
where
$\alpha$ is the fine-structure constant,
$\hbar$ is the plank constant,
and $c$ denotes the velocity of light.
It is shown that, both the critical photon energy and the radiation power can be enhanced by
increasing the values of electron energy, transverse velocity, and transverse oscillation frequency.
Laser wake field in  under-dense plasma is a promising medium for compact high-brightness source of keV x-rays \cite{puk_04a,puk_04b,kneip_np}.
State-of-the-art laser plasma electron accelerators can now accelerate electrons to
GeV energies in centi-metres \cite{gev}. 
However, it is very difficult to increase the energy more than one order of magnitude.
Fortunately, there are still some ways to increase the other two values.
The transverse betatron velocity can be enhanced more than one order of magnitude
by resonance between the electron betatron motion and the
laser pulse.
By irradiating a petawatt laser pulse on a gas target, in the direct laser acceleration dominated regime \cite{puk_dla, gahn}, high-brightness synchrotron X-ray can be generated \cite{kneip_prl}.
In laser wake field,
the betatron oscillation amplitude of GeV electrons can be dramatically enhanced
when resonance occur. 
By interacting the relativistic electrons with the rear of the driven laser pulse,
$10^8$ gamma-ray photons with spectra peaking between
$20$ and $150keV$
have been observed in experiment \cite{cipi_np}.
On the other hand,
by colliding high energy electrons with a laser pulse,
the transverse oscillation frequency can be an order of magnitude as the laser frequency,
which is usually two orders of magnitude higher than the betatron frequency in the wake field.
With the combination of a laser-wake-field accelerator and a plasma mirror, $10^8$ X-ray photons with photon energy ranging from $50keV$ to $200keV$ have been generated in experiment \cite{phuoc}.
With further optimizing, $10^7$ $MeV$ gamma-rays have
been emitted \cite{chen}.

\begin{figure}[h]
\centering
\includegraphics[width=0.5\textwidth]{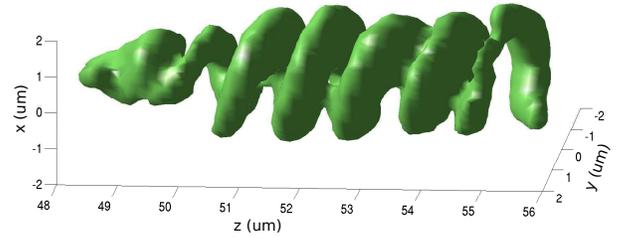}
\caption{(color online).
Isosurface plot of
electron energy density distribution with isosurface value $190n_cm_ec^2$ at time
$t=233fs$.}
\label{fig1}
\end{figure}

In this letter, we investigate
betatron radiation of electrons by propagating a ultra-intense ultra-short
laser pulse  in near critical density plasma.
We found that, both the transverse velocity $v_{\perp}$
and the betatron frequency $\omega_{\beta}$ can be enhanced dramatically.
In this condition, when the transverse betatron frequency
is close to the laser frequency in the electron frame,
relativistic electrons can undergo acceleration
and betatron oscillation simultaneously, and then a helical electron
beam can be generated \cite{smra}, as illustrated in Fig. \ref{fig1},
by propagating a 7.4J laser pulse in a near critical density plasma.
%
The relativistic electrons experience transverse oscillations with very high energy
and very high frequency, can emit high energy photons along electron motion direction.
In simulation,
8.3mJ electromagnetic radiation with
critical photon energy $E_c \sim 1.17 MeV$ is emitted.
Simulation results at different laser plasma parameters
show that,
$E_c$ can
increase with the initial laser energy $W_I$  as
$E_c \propto W_I^{1.5}$, and meanwhile the photon number $N_{\gamma}$ can be proportional
to $W_I$.

Here we normalized the  betatron oscillation frequency and transverse velocity by
$\nu=\omega_{\beta}/\omega_0$, and $\beta=v_{\perp}/c$, where
$\omega_0$ is the initial incident laser frequency.
According to the self-matching resonance acceleration regime \cite{smra},
for a resonance electron, we have
$\beta=\sqrt{\nu/2}$, and
 $\nu=1-v_z/v_{ph}$,
where $v_z$ is the electron velocity along laser propagation direction, 
$v_{ph}=\omega_0/k$ is the phase velocity of the laser pulse, and
 $k$ is the wave number which satisfies
$\omega_0^2=\omega_p^2+c^2k^2$.
The relativistic self-transparent plasma frequency $\omega_p$ can be written as
$\omega_p=\sqrt{4\pi e^2 n_e^2/a m_e}$,
where $a=e E_L/m_e c \omega_0^2$ 
is the normalized vector potential for a laser pulse with electric
field $E_L$ and laser frequency $\omega_0$,
 and $n_e$ is
the density of electron beam in the center of the laser channel.
The betatron frequency under azimuthal quasi-static transverse magnetic field
$B_{\theta}$ is \cite{smra}
$\omega_{\beta}=\sqrt{(e v_z/\gamma m_e)( \partial B_{\theta}/\partial r)}
=\sqrt{\mu_0 n_e e^2 v_z^2/\gamma m_e}$.
Then the maximum value of $\gamma$ accelerated by resonance
is $\gamma_{r}=\mu_0 n_e e^2 v_z^2/\omega_{\beta}^2 m_e$.
At the limit of $n_e/a \ll 1$ and $v_z \to c$,
one can get
\begin{equation}
\nu=\frac{n_e}{2a n_c}, \quad \beta=\frac{1}{2}\sqrt{\frac{n_e}{an_c}},
\quad \gamma_r=\frac{4 a^2 n_c}{n_e}.
\label{eq2}
\end{equation}
%
%
Then we can get
\begin{equation}
E_c \sim
16 \hbar \omega_0 a^{9/2}\left(n_e/n_c\right)^{-3/2}.
\label{eq3}
\end{equation}
It is appropriate to  assume that
every one electron experience one whole period to radiate.
Then the radiation energy per electron become
$w_r=P\times 2\pi/\omega_{\beta}=4\pi \alpha E_c \gamma_r \beta/3$.
The total energy of the betatron electrons can be written as
$W_{ele}= N_{\beta} \gamma_r m_e c^2,$
where $N_{\beta}$
denotes the total number of  betatron resonance electrons.
Then we can get the total radiation energy
\begin{equation}
	W_r=N_{\beta} w_r=\frac{4 \pi \alpha \beta}{3}  \frac{W_{ele}}{ m_e c^2}E_c,
\label{eq4}
\end{equation}
and the number of radiation photons with photon energy around $E_c$
\begin{equation}
	N_{\gamma}= W_r/E_c= \frac{4 \pi \alpha \beta}{3}  \frac{W_{ele}}{ m_e c^2}, 
\label{eq5}
\end{equation}
Further more, we can investigate the angle distribution of the radiation.
The peak of the angular distribution is at
\begin{equation}
	\theta_p \sim \arctan \beta \sim \beta, 
\end{equation}
and the divergence angle (full angle) is \cite{smra}
\begin{equation}
	\Delta \theta 
\sim \frac{\beta a}{\pi (R/\lambda)\left[ (B_{Sz}/B_0)^2+2(B_{Sz}/B_0) \right]},
\end{equation}
where $B_{Sz}$ denotes the axial magnetic field, and $R$ is the spot size of the field.

Now we present the details of the 3D simulations.
In our condition, the electromagnetic radiation is dominated by
synchrotron-like radiation regime.
When the pair generation can be ignored, and radiation coherence
is neglected,  the synchrotron-like radiation can be evaluated by calculating
the Lorentz-Abraham-Dirac equation.
However, the equation is very difficult to solve.
There are many modified methods to simplify the calculation \cite{nau_pop, chen_min}.
Here we extended a fully relativistic three-dimensional (3D) particle-in-cell (PIC) code
(KLAP) \cite{klap1,klap2}
by using the calculation method in Ref. \cite{nau_pop},
in which the radiation process and the recoil force are both considered consistently.
A 
circularly polarized (CP) laser pulse,
with
central wavelength $\lambda_0=1~ \mathrm{\mu m}$, wave period
$T_0=\lambda_0/c$, rising time $2T_0$, duration time $15 T_0$, ramping time $2T_0$,
and  a Gaussian transverse (X,Y) envelope
$a=a_0 \exp \left( -r^2/\sigma^2 \right)$,
here $\sigma=3\mu m$,
$a_0=13$ corresponding to a peak laser intensity $I=4.6\times
10^{20}~ \mathrm{W/cm^2}$,
is normally
incident from the left boundary ($z=0$) of a $100\times 12\times 12
~ \mathrm{\mu m^3}$ simulation box with a grid of $1200\times 144 \times 144$
cells. A near-critical density plasma target consisting of
electrons and protons is located in $6~ \mathrm{\mu m}<z<97~ \mathrm{\mu m}$.
In the laser propagation direction,  the plasma density rises linearly from $0$ to
$n_0=0.8 n_c$ in a distance of $5~ \mathrm{\mu m}$, and then remains constant,
where $n_c=m_e \omega_0^2  \epsilon_0 / e^2$ is the critical plasma density,
$m_e$ is the electron mass,
and $\epsilon_0$ is  the vacuum permittivity. In the radial direction, the density is uniform.
The number of super-particles used in the simulation  is about $1.8\times 10^8$ for each species
(8 particles per cell for each species corresponds to $n_0$).
An initial electron
temperature $T_e$ of $150 ~ ~ \mathrm{keV}$  is used to resolve the initial
Debye length ( $T_i= 10  ~ ~ \mathrm{eV}$ initially).

\begin{figure}[h]
\centering
\includegraphics[width=0.5\textwidth]{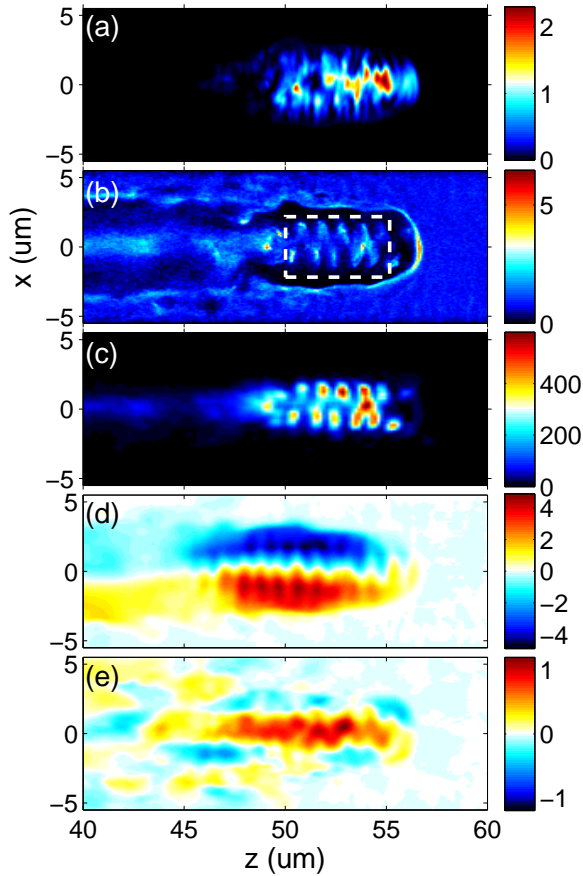}
\caption{(color online).
Longitudinal (Z, X) cuts along the laser pulse axis at $t=70T_0$,
(a), instantaneous laser intensity distribution $I$,
normalized by the initial intensity $I_0=4.6\times 10^{20}~ \mathrm{W/cm}^2$;
(b), electron density distribution $n_e$, normalized by the critical density $n_c$;
(c), electron energy density distribution, normalized by $n_c m_e c^2$;
(d)(e), self-generated quasi-static azimuthal and axial magnetic fields $B_{S\theta}$ and $B_{Sz}$,
averaged over $4$ laser periods, normalized by $m_e \omega_0/e$.}
\label{fig2}
\end{figure}

Figure \ref{fig2} presents snapshots of simulation results
at $t=70T_0$.
After a stage of filamentary and self-channelling, 
about $3/4$ of the laser energy has been exhausted by the plasma.
The laser pulse is slightly self-focused, and the laser intensity is close to the initial intensity, i.e., $a\sim a_0$, as
shown in Fig. \ref{fig2}(a).
Both electrons and ions are expelled by the self-focused laser pulse,
and a laser channel is formed.
A strong current of relativistic electrons is driven by the laser pulse in
the direction of light propagation, and confined in the laser channel.
A helical high density electron beam is formed
 in the center of the laser channel.
In the longitudinal (Z, X) cut of the electron density,
the helical beam shows a zigzag profile,
as shown in Fig. \ref{fig2}(b), labeled by a white dashed box.
The density of the beam is about $n_e \sim 2n_0$.
Then according to Eq.(\ref{eq2},\ref{eq3}), we can get that,
\begin{equation}
 \nu=0.062, \quad \beta=0.175, \quad \gamma=422, \quad E_c=1MeV.
\label{eq8}
\end{equation}
The energy density distribution is shown in Fig. \ref{fig2}(c).
It is shown that, most of the electron energy is localized in the beam
in the selected box.
The total energy of the electrons in the selected box is $0.9J$,
which is $12\%$ of the initial laser energy.
Then we can get
\begin{equation}
W_r=9mJ, \quad N_{\gamma}= 6 \times 10^{10},
\label{eq9}
\end{equation}
according to Eq. (\ref{eq4},\ref{eq5}).
The isosurface of the energy density with isosurface value $190n_cm_ec^2$ in
 3D  is shown in Fig.\ref{fig1}, which shows a helical structure clearly.
A strong quasi-static
azimuthal magnetic field up to 0.5GG is generated by the strong electron current,
as shown in Fig. \ref{fig2}(d).
Meanwhile, a strong axial magnetic field up to 0.12GG, with spot size $R \sim 1\mu m$
is generated, as shown in Fig. \ref{fig2}(e).
Then we can get $\Delta \theta \sim 0.18 rad $.
In this condition, electron acceleration is dominated by the self-matching resonance
acceleration regime \cite{smra}.
The accelerated relativistic electrons are executing collective
circularly betatron motion.
%

\begin{figure}[h]
\centering
\includegraphics[width=0.45\textwidth]{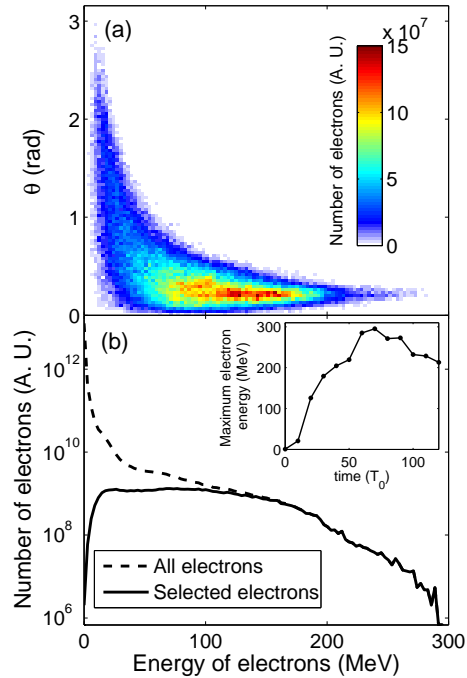}
\caption{(color online). 
(a) Energy angular distribution of electrons in the selected box in Fig. 2(b) at $t = 70 T_0$.
(b) Energy spectra of electrons in the selected box (solid line),
and all electrons (dashed line) at time $t=70T_0$.
Inset figure shows time evolution of the maximum
electron energy.
}
\label{fig3}
\end{figure}

%

The spectra property of electrons in the selected box at $t=70T_0$ is shown in Fig. \ref{fig3}.
The energy angular distribution shows that,
most of the high energy electrons is distributed at a same angle of
$\theta\sim 0.18 rad $, with a divergence angle (full angle) of $\Delta \theta \sim 0.15 rad $,
although the energy is ranging from $50MeV$ to $290MeV$, as plotted in Fig. \ref{fig3}(a).
This means that the high energy electrons are executing a collective circularly betatron motion,
with a transverse velocity $\beta =0.18$, and a Lorentz factor $\gamma$ ranging from
$100$ to $550$.
The simulation results
 coincide with the theoretical estimation.
%
The energy spectrum of electrons in the selected box
 exhibits a plateau profile distribution,
as shown in Fig. \ref{fig3}(b) by a solid line.
The inset figure plots time evolution of the maximum energy of electrons.
The electron energy increases dramatically at the begin, then reaches the maximum
value $300MeV$ at $t=70T_0$, and then decreases slowly, since the driven laser pulse
is exhausting.
The energy spectrum of all electrons is
 shown in Fig. \ref{fig3}(b) by a dashed line.
It is shown that, most of the high energy electrons are included in
the selected box in Fig. \ref{fig2}(b).

\begin{figure}[h]
\centering
\includegraphics[width=0.5\textwidth]{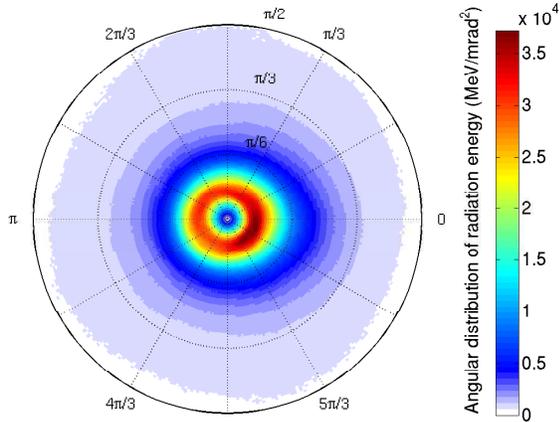}
\caption{(color online).
Angular distribution of radiation energy with photon energies above $100keV$.
The radial coordinate and the angular coordinate,
labels the the polar angle $\theta$ and  the azimuthal angle $\phi$
along the laser propagation direction, respectively.
}
\label{fig4}
\end{figure}

The angular distribution of the final radiation with photon
energies above $100keV$ is shown in Figure 3(a).
The distribution is approximately azimuthal symmetric
about the laser propagation direction,
and  most of the radiation energy is
distributed in a polar angle ranging from $0.12 rad$ to $0.35 rad$,
with a peak value $3.7\times 10^4 MeV /mrad^2$ at about  $0.2 rad$.
The final radiation distribution is a result of the energy angular distribution of high energy electrons, and confirms that most of the high energy electrons are executing
collective circularly betatron motion.
The total radiation energy calculated by integrating all the angles is about $8.3mJ$, which
is $0.1\%$ of the incident laser energy.
The corresponding photon number is $6.6\times 10^{10}$.
The simulation results close to above theoretical estimation.
It is noticed that,
most of the radiation energy emitted with a finite polar angle, rather than that in most cases
along the laser propagation direction.
This is because that
the synchrotron-like radiation of relativistic electrons is
emitted almost along the electron motion direction, and
the resonance electrons have a large  transverse velocity.
The duration time of the gamma-ray pulse is close to the length of the electron beam,
is about $17fs$. 
Since the radius of the radiation source is less than $1\mu m$, then
we can get the brightness of the gamma-ray emission with energies above $0.1E_c$
is $1.5\times 10^{22} \rm{ photons/s/mm^2/mrad^2/0.1\% bandwidth}$.

\begin{figure}[t]
\centering
\includegraphics[width=0.45\textwidth]{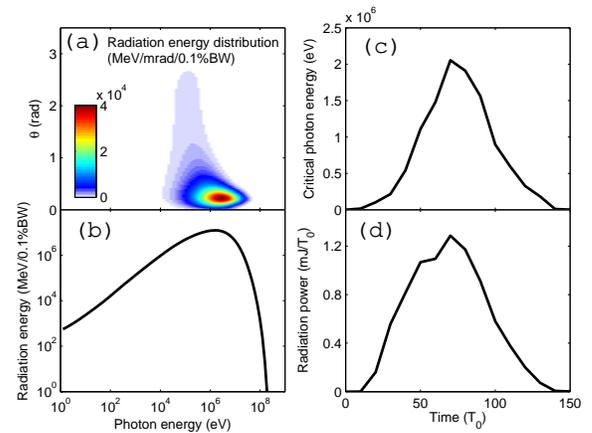}
\caption{(color online).
(a) Polar-angularly and spectrally resolved radiation energy.
(b) Radiation spectrum (radiation energy per $0.1\%$ band width (BW)).
(c)(d), Time evolution of critical photon energy, and total radiation power,
respectively.}
\label{fig5}
\end{figure}

More details of the radiation is shown in Fig. \ref{fig5}.
Since the radiation is azimuthal symmetric, we can plot the
polar-angularly and spectrally resolved radiation energy,
as shown in Fig. \ref{fig5}(a).
It is shown
that, most of the radiation energy is distributed at a
peak angle $\sim 0.2rad$, and a
divergence angle (full angle) about $\sim 0.2rad$,
with photon
energy ranging from $100keV$ to $20MeV$.
The radiation energy spectrum by integrating the polar angle is
shown in Fig. \ref{fig5}(b).
It is shown that, The peak of the spectrum is located at $1.3MeV$.
Since the spectrum is synchrotron-like, we can define a critical
photon energy, divided by which the integration of the two parts are equal.
Here the critical photon energy is $1.2MeV$
close to the peak value,
and agree well with above theoretical estimation.
Fig. \ref{fig5}(c)(d) show time evolution of the critical photon energy and
the radiation power, respectively.
They are calculated by analyzing the radiation every per $10$ laser periods.
It is shown that, the critical photon energy and
the radiation power show similar evolution  in time.
After a fast increasing,
both the critical photon energy and the radiation power  reach peak values
At $t=70T_0$.

\begin{figure}[t]
\centering
\includegraphics[width=0.45\textwidth]{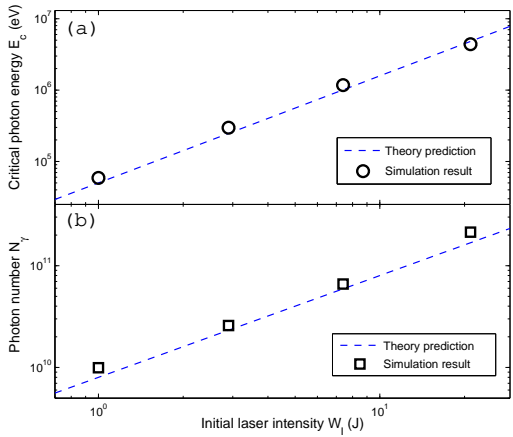}
\caption{(color online). Variation of (a) critical photon energy, (b) total radiation energy,
with the incident laser energy $W_I$, by keeping
$l_s=\sqrt{a n_c/n_e}$ fixed.
}
\label{fig6}
\end{figure}

Above investigation can be extended to a large range of laser energies.
%
We simulated different laser plasma parameters, by keeping
 the dimensionless plasma skin length
$l_s=\sqrt{a n_c/n_e}$ fixed,
with initial laser energy ranging from $1J $ to $21J$.
We found that, the laser plasma interactions exhibit a
scaling property on $l_s$, especially, by keeping $l_s$ fixed,
the values of $n_e/n_0$ and $W_I/W_{ele}$
nearly keep constant.
Many other works also show that there is a scaling on $l_s$ \cite{scale_puk,wang_scale}.
Since the energy of the laser pulse
$W_I=2\pi \sigma^2 T_L I =2\pi \sigma^2 T_L  I_1 a^2$,
where  
$T_L=17T_0$ is the effective laser duration time,
and $I_1=1.37\times 10^{18}W/cm^2$ is the laser intensity when $a=1$,
then we can get the critical photon energy as a power function of the initial laser energy as
\begin{equation}
	E_c =
16 \hbar \omega_0  l_s^{3} \left(\frac{W_I}{2 \pi \sigma^2 T_L I_1}\right)^{3/2}
= 5\times 10^4 W_I[J]^{1.5} (eV).
\label{ecn}
\end{equation}
And the number of the gamma photons is proportional to $W_I$,
\begin{equation}
	N_{\gamma}=\frac{4\pi \alpha \beta}{3 m_e c^2} \frac{W_{ele}}{W_I} W_I
= 8\times 10^9 W_I[J].
\label{ngn}
\end{equation}
The simulation results of the critical photon energy $E_c$
and the photon number with photon energies above $0.1E_c$ are
shown in Fig. 5(a),(b), respectively.
The dashed lines are the theoretical estimation of Eq. (\ref{ecn}),(\ref{ngn}).
The simulation results agree well with the theoretical estimation.
The critical photon energy
is increasing with the initial laser energy much faster than a linear relation,
which is the upper limit of the X-ray radiation in under-dense plasma \cite{rmp_x-ray}. It is noticed
that the critical photon energy, gamma photon number and radiation spectrum 
 are similar in case of Linear Polarized laser pulse, only the Angular distribution of radiation
energy is little different.
%

In conclusion, we have investigated electromagnetic emission
by propagating an $7.4J$ ultra-intense ultra-short laser pulse in a near critical density plasma.
$6.6\times 10^{10}$ gamma-ray photons with critical photon energy $1MeV$ are emitted when electrons experience betatron resonance acceleration.
With the initial incident laser energy $W_I$ increasing, the
 critical photon energy $E_c$ and the photon number $N_{\gamma}$
 increase as
$E_c \propto W_I^{1.5}$, and $N_{\gamma} \propto W_I$, respectively.

This work was supported by National Basic Research Program of China
(Grant No. 2013CBA01502),National Natural Science Foundation of
China (Grant Nos. 11025523,10935002,10835003,J1103206) and National
Grand Instrument Projetc(2012YQ030142).

\end{document}